\documentstyle[rotating,epsfig,amsmath,amssymb]{elsart}

\begin{document}

\begin{frontmatter}

\title{The Time-of-Flight Technique \\ for the HERMES Experiment}

\author[yer]{A.~Airapetian}, 
\author[yer]{N.~Akopov}, 
\author[rom,yer]{M.~Amarian}, 
\author[fra,yer]{H.~Avakian}, 
\author[yer]{A.~Avetissian}, 
\author[fra,yer]{E.~Avetisyan}, 
\author[cal]{B.W.~Filippone}, 
\author[gla]{R.~Kaiser}, 
\author[yer]{H.~Zohrabian}, 

\hyphenation{HERMES}

\address[cal]{W.K. Kellogg Radiation Lab, 
California Institute of Technology,\\ 
Pasadena, California 91125, USA}
\address[fra]{INFN, Laboratori Nazionali di
Frascati, 00044 Frascati, Italy}
\address[gla]{University of Glasgow, Dept. of Physics and Astronomy,\\
Glasgow G12 8QQ, Scotland, U.K.}
\address[rom]{INFN, Sezione Sanita, 
00161 Roma, Italy}
\address[yer]{Yerevan Physics Institute, 375036 Yerevan, Armenia}
%\address[zeu]{University of Glasgow,
%Dept. of Physics and Astronomy,
%Glasgow G12 8QQ
%United Kingdom}
\newpage

\begin{abstract}
This paper describes the use of the time-of-flight (TOF) technique 
as a particle identification method for the HERMES experiment.
The time-of-flight is measured by two $1\times 4 m^2$ scintillation 
hodoscopes 
that initially were designed for the first-level trigger only. 
However, the suitable time structure of the HERA electron beam allows
an extension of their functions to also measure the TOF for low momentum
hadron identification. Using only these conventional hodoscopes, good 
particle identification was achieved for
protons and pions in the momentum range up to 2.9~GeV/c and for
kaons up to 1.5~GeV/c.\\
{PACS number 29.30.-h, 29.40.-n, 29.40.Gx, 29.40.Mc}
\end{abstract}
\end{frontmatter}

\twocolumn

\section{Introduction}
The HERMES experiment at DESY~\cite{specpaper} is a second generation 
polarised deep-inelastic scattering experiment to study the spin 
\linebreak structure 
of the nucleon. Several experiments over the last decade have provided   
accurate data on the polarisation asymmetry of the cross-section for 
{\it inclusive} scat\-tering where only the scattered lepton is detected.
Further knowledge of the origin of the nucleon's spin can be gained by studying
{\it semi-inclusive} processes involving the detection of hadrons in
coincidence with the scattered lepton. This increases the demands 
on hadron identification of the detection system.\\ 
The time-of-flight (TOF) particle identification method 
is a fast, inexpensive and efficient technique~\cite{atwood} for hadron 
identification. Its implementation at HERMES is possible because of 
a) the presence of two scintillator counter walls H1 and
H2 in the HERMES spectrometer, and b) the fine time structure of the
HERA electron beam (with bunch lengths of 27 ps and time between bunches
of 96 ns), which allows particles from different
bunches to stay completely separated in time.

\section{Counter Design}

\begin{figure}
%Fig.1
   \epsfig{file=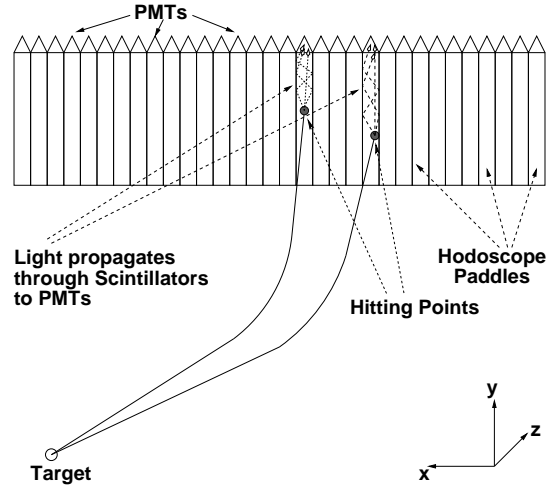,width=7.5cm,angle=0}
      \caption{
Schematic view of particle tracks and hodoscope paddles (upper part).
}
\label{hodoscope}
\end{figure}

HERMES is a forward spectrometer with a large dipole magnet, a set of tracking
detectors, and particle identification (PID) detectors consisting of 
a \v{C}erenkov detector, a Transition ~ Radiation ~ Detector (TRD), 
~a preshower detector and a 
calori-meter~\cite{specpaper}. In 1998 the threshold-\v{C}erenkov 
coun\-ter 
was
upgraded to a Ring
Imaging \v{C}erenkov
(RICH) 
detector~\cite{rich}.
A scintillator hodosco\-pe (H1) and a Pb-scin\-tillator preshower 
counter (H2) (Fig.~\ref{hodoscope}) provide trigger
signals and particle identification information. Both
counters are composed of vertical scintillator modules (42 each in the
upper and lower detectors), which are 1 cm thick and 9.3 cm x 91 cm in 
area.
The material for the modules is BC-412 from Bicron Co., a fast  
scintillator with large attenuation length (300-400 cm for scintillation 
light). The
scintillation light is detected by 5.2 cm diameter Thorn EMI 9954
photomultiplier tubes coupled via a light guide to the outer end of
each scintillator (away from the beam plane). The modules are staggered to
provide maximum efficiency with 2-3 mm of overlap between each unit.
Each hodoscope photomultiplier (PMT) signal is passively
split with one output going to a LeCroy 1881M ADC
and the  other going to a LeCroy 3420 Constant Fraction
Discriminator (CFD). The individual CFD outputs are fed to LeCroy 1875A
time to digital converters (TDC), which measure the time-of-flight
using the HERA-clock as a reference signal. This
signal corresponds to the moment when the HERA bunch cros{-}ses the center
of 
the target. When some interesting event configuration is recognized by the 
HERMES trigger logic, the HERA-clock signal is enabled to start the TDC 
modules  in common-start mode, 
with the STOP signals
coming from each 
scintillator. 
The ~ time ~ base ~ for ~ the TDC
is 50 ps/channel. 

\section{Calibration procedure}

\begin{figure}[h]
%Fig.2 
\epsfig{file=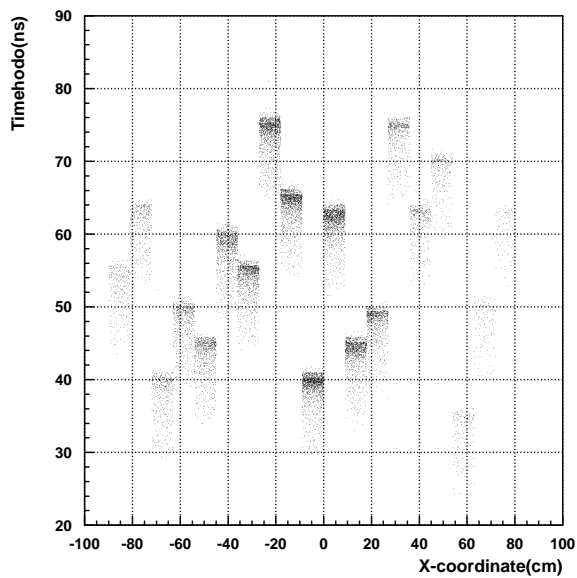,width=7.5cm}
\caption{Non-
equalized time dependence over X-coordinate}  
\label{xcoor}
\end{figure}

\begin{figure}
%Fig.3
\epsfig{file=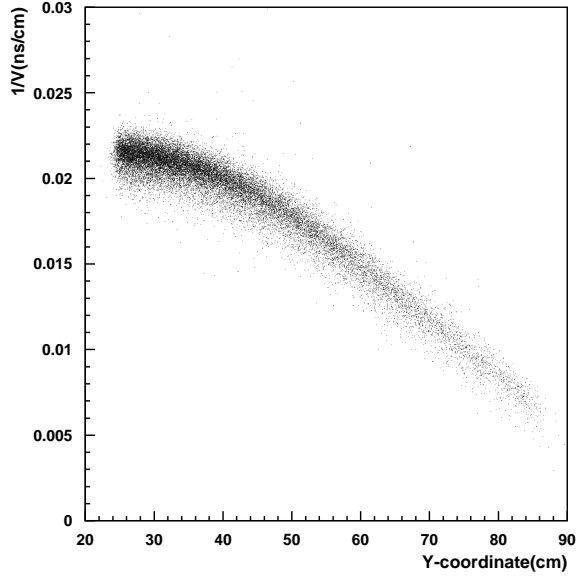,width=7.5cm}
\caption{$1/v_e$ non-equalized dependence on the $Y$-coordinate for
electrons (single paddle).}
\label{vvsY}
\end{figure}

\begin{figure}[h]
%Fig.4
\epsfig{file=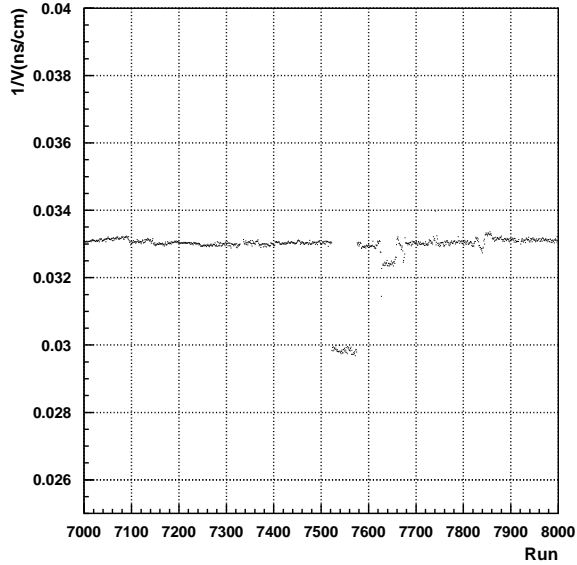,width=7.5cm}
\caption{$Y$ corrected $1/v_{e}$  dependence on run numbers (for electrons).}
\label{timedep}
\end{figure}

\begin{figure}
%Fig.5
  \epsfig{file=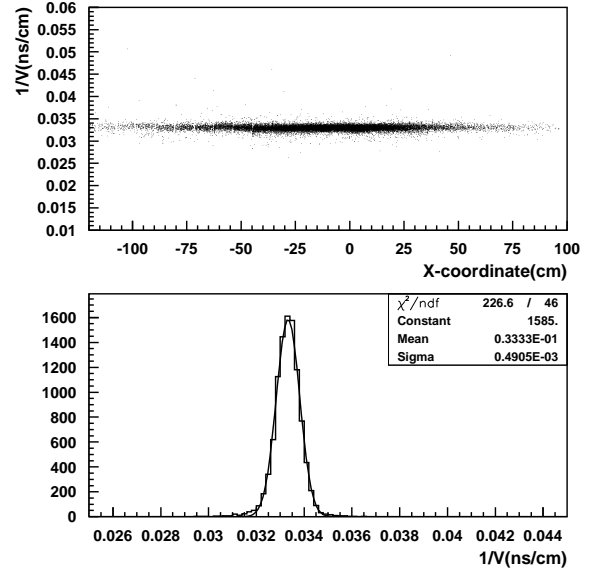,width=7.5cm}
\caption{$1/v_{e}$ for electrons in hodoscope H1 
(equalized and run-corrected).}
\label{finalv}
\end{figure}

The calibration procedure is based on the fact
that electrons above 10 MeV are moving at essentially the speed of light. 
Any measured deviations from this must be artifacts of
the experiment that should be corrected.\\
In Fig.~\ref{xcoor} the time distribution for electrons over the paddles 
is shown for part of one detector. The electron sample is selected by a 
combination of cuts on TRD, \v{C}erenkov, Preshower and Calorimeter.
The time distribution is spread within each paddle as well as 
between paddles. The differences between the paddles (due to different
cable lengths, etc) were removed by
setting the average speed of the electrons to the speed of light 
for each paddle individually. 
A single-counter time response is described by
\begin{equation}
        t= t_{pf}+t_{lf}+t_{0}
\label{teq}
\end{equation}
where
$t_{pf}$ is the particle time-of-flight from the interaction point
to the scintillator \linebreak plane,
$t_{lf}$ is the time needed for the light created in the scintillator
to reach the PMT,
$t_{0}$ is the constant time offset specific to each paddle, arising from 
the PMT response time, cable delays, TDC calibration intercept, etc.
The path length $l_H$ of each track to hodoscope H was calculated using 
the reconstructed track parameters, along one straight line segment from 
the interaction point in the target to the mid-plane of the spectrometer 
magnet, and continuing along another straight line from there to the plane 
of the hodoscope H. This approximation is adequate because the magnet bend 
angle is less than  $\pi/20$.
The calibration for each paddle n of hodoscope H was done with electron 
tracks by fitting the y-distribution of $1/v=t/l_H$ shown in  
Fig.~\ref{vvsY} to the known value of $1/c$ using a fourth-order 
correction polynomial $F_n(y)$:\\ 
$F_n(y) = 1/c - t/l_H$.\\
The constant term in the polynomial incorporates $t_0$ in Eq.~1 for each 
paddle.
A similar method has been used in \cite{synodinos} where a more
detailed explanation for each coefficient of the polynomial is given.
Occasional shifts in the HERA-clock signal derived from the electron beam 
accelerator system, shown in Fig.~\ref{timedep}, were compensated by 
re-fitting the constant term for each run of approximately 10 minutes. 
The final corrected distribution of $1/v_e = t/l_H + F_n(y)$ for electrons 
is shown in Fig.~\ref{finalv}.
The  resolution  extracted from a Gaussian fit corresponds to
$\sigma$=0.49 ns.

\section{Hadron Identification via \newline Time-of-Flight}
\begin{figure}
%Fig.6
  \epsfig{file=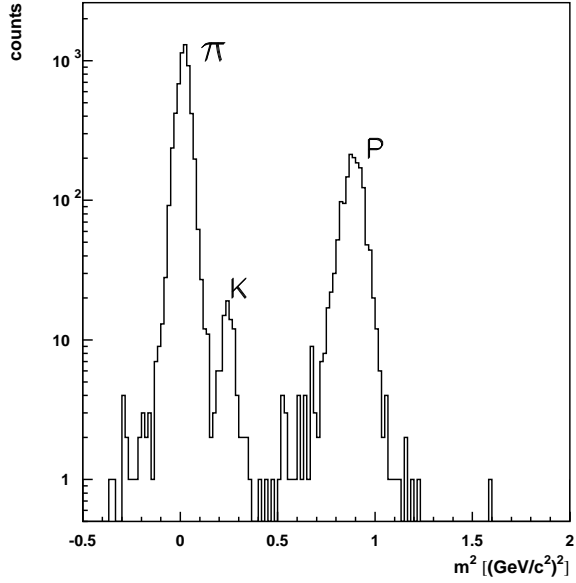,width=7.5cm}
\caption{$m^2$ distribution for hodoscope H1 upper detector), 
$0.6<p<1.1$ GeV/c. Log scale is chosen due to significant difference 
between different particles fluxes.}
\label{m21}
\end{figure}

\begin{figure}
%Fig.7
  \epsfig{file=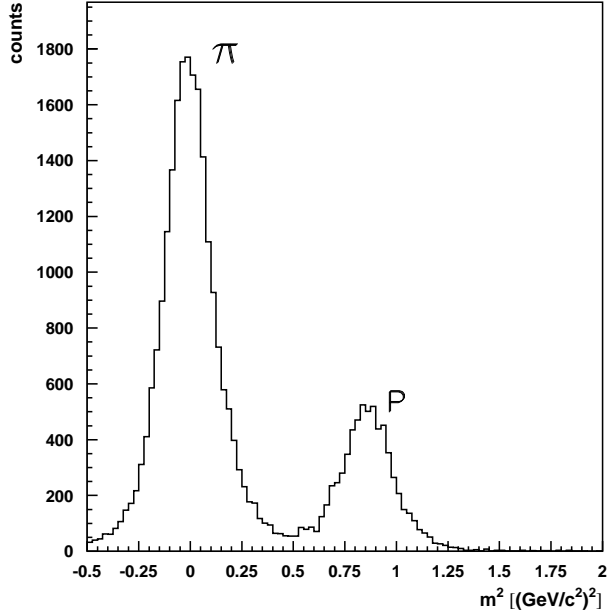,width=3.125in}
\caption{$m^2$ distribution for hodoscope H1 (upper detector), 
$1.5<p<2.0$ GeV/c}
\label{m22}
\end{figure}

\begin{figure}
%Fig.8
  \epsfig{file=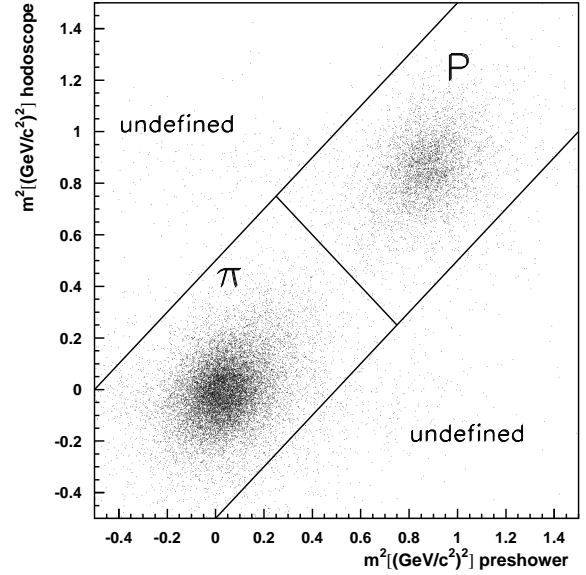,width=7.5cm}
\caption{Two-dimensional distribution of squared mass from Hodoscope(H1) 
and Preshower(H2) for
the momentum region $1.5<P<2.0$ GeV/c.}
\label{2d}
\end{figure}

From the relativistic momentum of the particle
\begin{equation}
 p = m\cdot\beta/\sqrt{(1-\beta^2)},
\end{equation}

the mass $m$ is extracted using the speed $\beta=v/c$ obtained through the hodoscope timing:
\begin{equation}
 m^2=p^2 (\frac{1}{\beta^2} - 1).
\end{equation}
The squared mass of the particle was chosen as the parameter for
the identification. To check the algorithm and calibration,  the proton 
and kaon  squared masses have been extracted giving values of 
$m_{p}^2=0.88$ (GeV/c$^2$)$^2$ and $m_{K}^2=0.25$ (GeV/c$^2$)$^2$,
respectively, 
which agrees well with the expected values (Fig.~\ref{m21}).  
At  momenta below $2$ GeV/c the kaon flux is two orders of  magnitude 
smaller 
than the pion flux, 
so that no separation is possible via TOF for momenta above $1.5 $ GeV/c. 
Therefore  in the momentum range $1.5 < p < 2.0$ GeV/c, kaons are included 
in the pion 
spec\-trum as a negligible contamination ($< 1$ \%). At higher momenta, a 
RICH detector takes over.\\
Fig.~\ref{m22} shows the squared hadron mass distribution for the momentum 
region $1.5<p<2.0$ GeV/c. The good separation between protons
and pions is clearly visible. Using TOF information from only one detector 
(e.g. hodoscope H1), ~ the upper bound for the separation
is about 2.4 GeV/c.\\
The availability of two independent detectors ~ H1 and ~ H2
 suggests the use of ~ two-dimensional distributions, as shown in
 Fig.~\ref{2d}. 
Independent constraints on the TOF values from both hodoscopes is the best 
strategy to
minimize the contamination. However, this is not the best way to maximize
the efficiency while maintaining a low contamination. In the case of two
detectors a linear {\it 'valley cut'} in the plane of the two detector
responses gives an improvement~\cite{ralf}. 
It is possible to apply the combined constraint on the sum $m^2_{H1} + 
m^2_{H2}$.
%\begin{figure}
%%Fig.9
%  \epsfig{file=projection.eps,width=7.5cm}
%\caption{Valley cut in the detector 1,2 plane and projection onto the 
%axis
%perpendicular to the cut.}
%\label{vcut}                                
%\end{figure}
This evaluation provides an improvement in the separation between 
protons and pions
- especially for higher momenta - and therefore  extends the
momentum region for hadron identification with high efficiency and low
contamination. The identification of kaons is limited to $p<1.5$ GeV/c, 
due to the very low kaon flux (see Fig.~\ref{m21}).
\begin{figure}
%Fig.10
  \epsfig{file=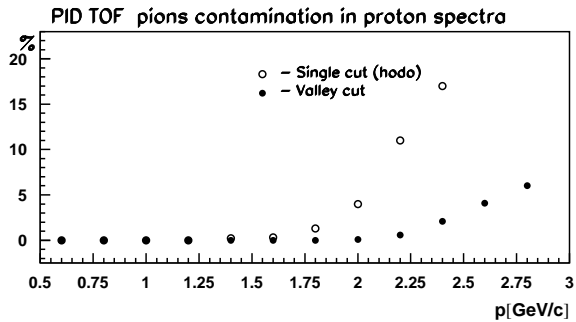,width=7.5cm}
\caption{
Pion contamination in the proton sample for the cases of 
individual constraints on one of the two hodoscopes with that for the 
combined 
constraint.}
\label{contamination}
\end{figure}
The comparison of the pion contamination in the proton sample 
for individual constraints on the two hodoscopes with that for the 
combined constraint is presented
in Fig.~\ref{contamination}.\\
\begin{figure}
%Fig.11
  \epsfig{file=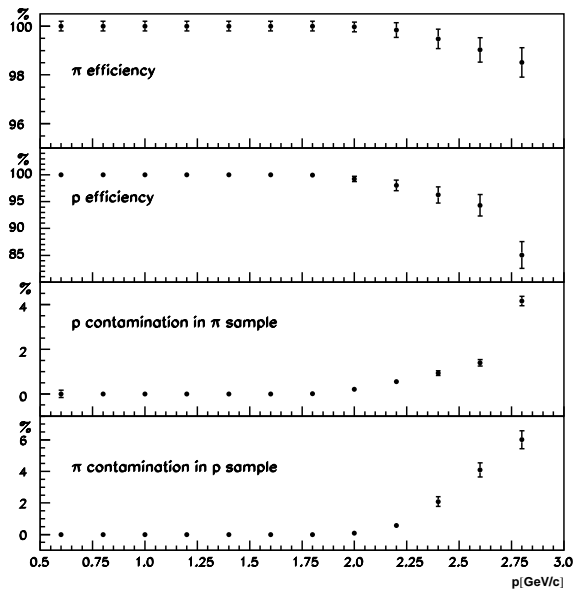,width=7.5cm}
\caption{
Proton and pion identification efficiency and contamination vs momenta
for the valley cut.
}
\label{effcon}
\end{figure}
The most important parameters for particle identification are 
the 
efficiency for identification of certain particle types and the
contamination from other types. Fig.~\ref{effcon} presents both parameters 
for proton and pion identification. 
The valley cut allows identification of protons and pions up to 2.9 GeV/c.
The resulting pion sample has an efficiency above 98\% and less
than 4\% proton contamination, while the protons have an efficiency
of more than 85\% and less than 6\% pion contamination in the
highest momentum bin. All analysis was done using semi-inclusive data 
samples 
from the year 1997 on polarized hydrogen target.
 
\section{Conclusion}
The TOF method, as presented in this paper, allows an extension of the 
momentum range for hadron identification in HERMES towards lower momenta. 
The application of this method provides more statistics for
hadrons detected within the HERMES acceptance and extends the
minimum value of $z=\frac{E_h}{\nu}$, where $E_h$ is the hadron energy and 
$\nu$ is the virtual photon energy.

\section{Acknowledgements}
We gratefully acknowledge the DESY management for its support and the DESY 
staff and the staffs of the collaborating institutions for their strong and
enthusiastic support.

\end{document}